# Vacuum Brazing

*S. Mathot*
CERN, Geneva, Switzerland

**Abstract**
First, we describe vacuum brazing and vacuum soldering. Then, we demonstrate how vacuum assists in reducing specific metal oxides, significantly enhancing the wetting of the braze material. Examples of metal-to-metal brazing and soldering are provided. The brazing of large accelerating cavities, specifically Radio Frequency Quadrupoles (RFQ), is then explained in more detail to present a comprehensive procedure tailored for high-precision, large-scale assemblies. Finally, we discuss the process of brazing ceramics using both techniques: with metallisation and with active brazing alloy.

## 1   Vacuum Brazing & Vacuum Soldering

Brazing is an ancient assembly technique. Objects that have been brazed more than 4,000 years ago have been found. Brazing and soldering use the same technique. To assemble components that are not necessarily made of the same material, a third material called the filler is used. This filler necessarily has a lower melting point. The components are heated together with the filler material placed in or close to the gaps between the components we want to join. The temperature is maintained just below the melting point of the filler material until all the components reach the same temperature. The temperature is then increased slightly above the melting point of the filler material, which flows by capillary action into the gaps between the components. This process is rapid, and the heating is stopped after a short time reducing the temperature below the melting point of the filler material, causing it to solidify immediately. The solidified filler material present in the gaps between the components ensures the assembly of the whole.

The difference between brazing and soldering is determined solely by the melting point of the filler material used: If the melting point is less than 450 °C, we called soldering and brazing above. Although the characteristics and applications of brazing and soldering are generally quite different, the principle is the same. Advantages and disadvantages are often valid for both:

– the assembly of different materials, metals or even ceramics is possible,

– the assembly can be of a high precision (for brazing mainly).

However:

– the mechanical resistance is generally less than for the base materials,

– the heat treatment generally deteriorates the mechanical properties of the assembled components.

Also, wetting by the filler material in liquid phase on the surface of the components is essential and is generally impossible in a normal atmosphere. The presence and the strong formation during the heat cycle of oxides at the surface of the components will indeed completely prevent the wetting of a liquid phase on a solid surface at high temperature.

One solution is to use a flux which is a chemical mixture, organic or not, that reduces and prevents the oxidation of the components and the filler material during the heating cycle performed in a normal atmosphere.

Another solution is to perform the heat cycle in vacuum (or sometime in a controlled atmosphere). This is known as vacuum brazing or vacuum soldering and has several advantages for the assembly of high precision, high cleanliness components used at CERN.

## 2    Oxide reduction in vacuum

It is possible to calculate that, at a given temperature, if the partial pressure of oxygen in the atmosphere is lower than a certain equilibrium value, the oxide that naturally forms on the surface of a metal is unstable and decomposes. For copper, the equilibrium value at 800 °C is $1.3 \times 10^{-6}$ Torr. In an all-metal vacuum furnace, such as those used at CERN, the oxygen partial pressure at 800 °C can be much lower than this value. This means that the surface of a pure copper component inside the furnace at this temperature is completely free of oxide. If a filler metal consisting of copper and silver (another metal with a similarly high equilibrium value) is used, the liquid phase will wet and flow perfectly over the surface of the pure copper component.

Other elements, such as aluminum, have a very low equilibrium value. For aluminum oxide at 800 °C, we calculate an equilibrium oxygen partial pressure as low as $2.5 \times 10^{-41}$ Torr. Such low oxygen partial pressure cannot be obtained in a vacuum furnace, so direct brazing of aluminum oxide (alumina) is impossible. As we will see later in the text, other tricks are needed for brazing alumina ceramics!

Even more elements have intermediate values. Chromium, for example, that is present on the surface of stainless steel. To easily braze stainless-steel in a vacuum furnace, we can depose a very thin layer of nickel on the surface of the stainless-steel components. Nickel has a higher equilibrium value than chromium, and vacuum brazing stainless steel components to copper components (a combination often used for accelerator elements) is perfect in a vacuum furnace using silver-copper-based filler material and nickel-plated stainless-steel.

The temperatures involved in soldering are necessarily low (below 450 °C), and the required equilibrium oxygen partial pressures are then very low and cannot be achieved in a vacuum furnace. Only silver oxide decomposes at the melting temperature of common solder alloys. In vacuum soldering, wetting is good on a silver layer and acceptable on well-cleaned copper. However, caution must be observed when using vacuum soldering on other metals, and the main advantage of this technique is the cleanliness of the assemblies obtained.

## 3    Vacuum brazing: Advantages & examples

In vacuum brazing, the wetting of the filler (or brazing alloy, in this case) is generally very good, and the liquid alloy flows easily between the assembled components by capillary action. Brazing is possible on large surfaces with a minimum of porosity. Vacuum brazing is therefore ideal when good thermal or electrical contacts are required.

After the thermal cycle in a vacuum furnace, the assemblies are perfectly clean and compatible with UHV applications, knowing that brazing alloys are always made of elements with a high purity and a low vapour pressure.

Unlike welding, the melting point of the base material is not achieved in brazing, and dissimilar materials can generally be assembled.

However, wetting by capillarity requires very narrow gaps. Depending on the brazing alloy used, the ideal gap may be as small as 25 μm (see Table 1).

**Table 1:** Recommended gap for some brazing alloys in vacuum brazing

| Brazing alloy | Gap (mm) | Ideal (mm) | Brazing Temp. (°C) |
|---|---|---|---|
| Cu | 0-0.05 | 0.025 | >1083 |
| Ag-Cu, Ag-Cu+Pd | 0-0.05 | 0.025 | 795-850 |
| Ag-Cu+In | 0.05-0.1 | 0.05 | 750-770 |
| Au-Cu | 0.05-0.1 | 0.05 | >920 |
| Ni-Cr | 0.03-0.15 | 0.05 | >1050 |

Low gaps mean high-precision machining, and the preparation of components for vacuum brazing assembly is often challenging. The tolerances are in many cases in the range of ± 0.01 mm, flatness 0.02 mm and finishing Ra 0.8 μm. However, the result is assemblies that can be produced with very high precision and very little deformation.

Table 1 indicates a gap of 0 mm for Cu or Ag-Cu brazing alloys. This is the 'zero-gap' technique, in which the brazing surfaces are in contact prior to brazing, with the brazing alloy flowing in between only by capillary action. This allows the greater mechanical precision in the assemblies.

Figure 1 shows examples of vacuum-brazed assemblies. We have chosen for the first examples copper/stainless steel junctions, which represent the majority of assemblies made by vacuum brazing at CERN. Picture A shows a small copper piece, measuring approximately 1 cm in length, brazed onto a nickel-plated stainless-steel dome. The silver-based alloy wets both the copper and stainless-steel surfaces very well. The brazing alloy forms a perfect neck at the junction between the two pieces. Picture B shows a large assembly: the copper ring has a diameter of about 400 mm and is brazed onto a thick stainless-steel disc. The homogeneous presence of the brazing alloy along the internal and external diameters is well visible.

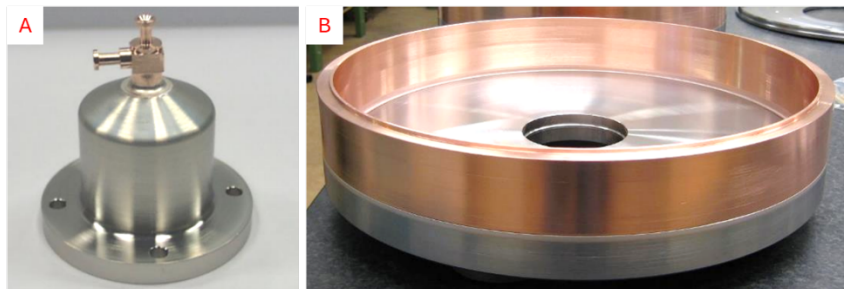

**Fig. 1:** Examples of copper/stainless steel vacuum brazing

Figures 2 and 3 illustrate a typical application of copper/stainless steel vacuum brazing: The heat exchanger tubes (HET). Approximately 1,700 copper tubes were required for the dipole and quadrupole cold masses of the LHC. Each tube is around 15 metres long and 58 millimetres in diameter, and a stainless-steel junction must be assembled at each end, see Ref. [1]. The most effective construction method is to vacuum braze a small copper sleeve to the end of each stainless-steel junction and weld the copper sleeve to the long tube using an electron beam.

Figure 2 shows the three HET models, with the stainless-steel junction (1), copper sleeve (2), electron beam welding (3) and tube (4) indicated. For technical and economic reasons, the tubes were made of OF copper, but the sleeve was made of Cu-OFE (grade UNS C10100). Using copper with a higher oxygen concentration than OFE copper is not recommended for vacuum brazing.

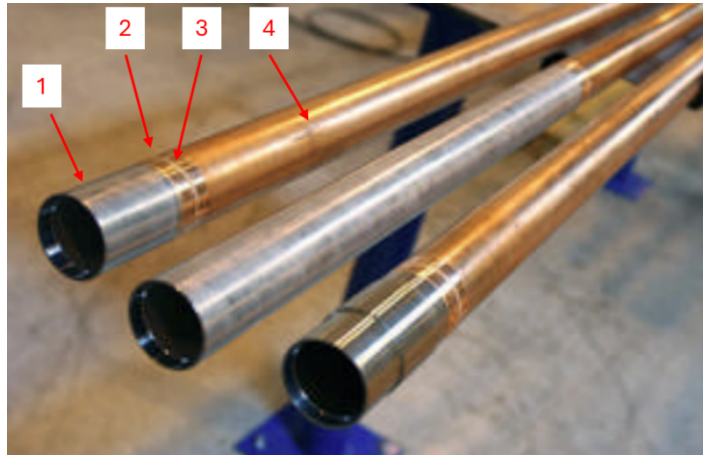

**Fig. 2:** Extremity of three versions of the HET

Figure 3 presents the brazing process in detail. Picture A shows the bimetallic junctions inside the vacuum furnace just before brazing. We see the support grid made of molybdenum (1), the ceramic spacer (2), the stainless-steel junction (3), the extremity of this junction having be nickel plated (4). The region where the plating is removed to reduce excessive flow of the brazing alloy (5) can also be seen, as well as the OFE copper sleeve (6), the ceramic spacer (7) and the stainless-steel cylinder (8) used to press on the assembly during the heat cycle.

Pictures B, C and D show the details of the junction. The ends of both, the copper sleeves and the stainless-steel junctions, are machined with precision. The copper sleeves are on the outside and are inserted into the stainless-steel junction over a length of 7.5 cm. The arrow in the picture B points to the groove machined in the stainless-steel junction into which the brazing alloy is placed. Dimensions are 0.8 x 0.8 mm, radius 0.4 mm and the brazing alloy made of silver, copper and palladium is in form of a metallic wire diameter 0.75 mm. The tolerances were +0.01 / +0.03 mm for the machining of the brazed diameter on the copper and 0 / -0.02 mm for the corresponding diameter on the stainless-steel. After brazing, we can see that the brazing alloy is completely melted, the groove is empty, the brazing alloy has flowed perfectly all along the 7.5 mm gap and is present on the internal (picture C) and external (picture D) diameters of the brazed junction.

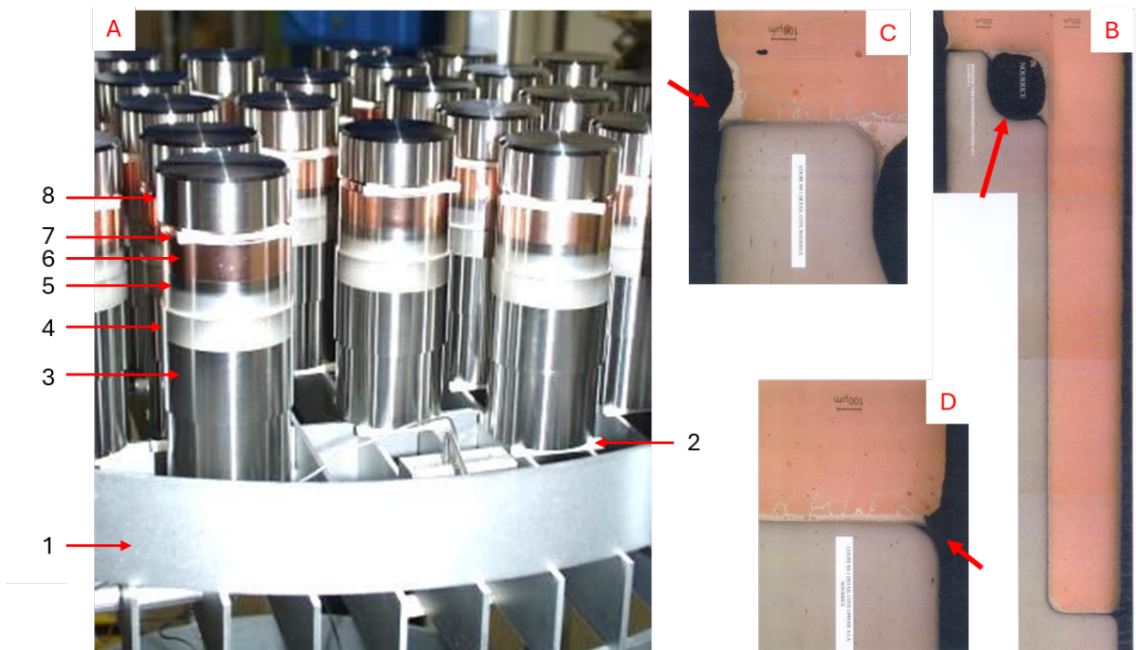

**Fig. 3:** Metallographies of a HET vacuum brazed junction.

Visual inspection of the assemblies after brazing is essential. The presence of the brazing alloy on these internal and external diameters (red arrows in pictures C and D) must be perfectly continuous, with no apparent gaps. This indicates that the alloy has filled the brazed gap and that there is no risk of a virtual leak with the brazing groove. If any gaps are visible, the assembly must be rejected. Otherwise, the assembly can be validated with a vacuum test.

Although welding is widely used for assembling stainless-steel components, vacuum brazing can sometimes be advantageous. This is the case, for example, in large-scale production of small pieces or for delicate assemblies with thin walls (honeycomb). Figure 3 provides two examples: A large electropolished hood (610 mm in diameter) and a grid, see picture A. The advantages for brazing were minimizing the deformations and achieving a perfectly clean assembly after brazing. Picture B shows the quality of the wetting of the nickel-based brazing alloy on the electropolished stainless-steel surfaces. The second example (Picture C) is the brazing of a thin stainless-steel foil (0.1 mm thick) (1) to close an opening made in the bottom wall of a rectangular box (2). In front of the opening, a very good flatness of the foil was required. With vacuum brazing we minimized the deformations, and we obtained a planarity as low as 30 microns, see Ref. [2].

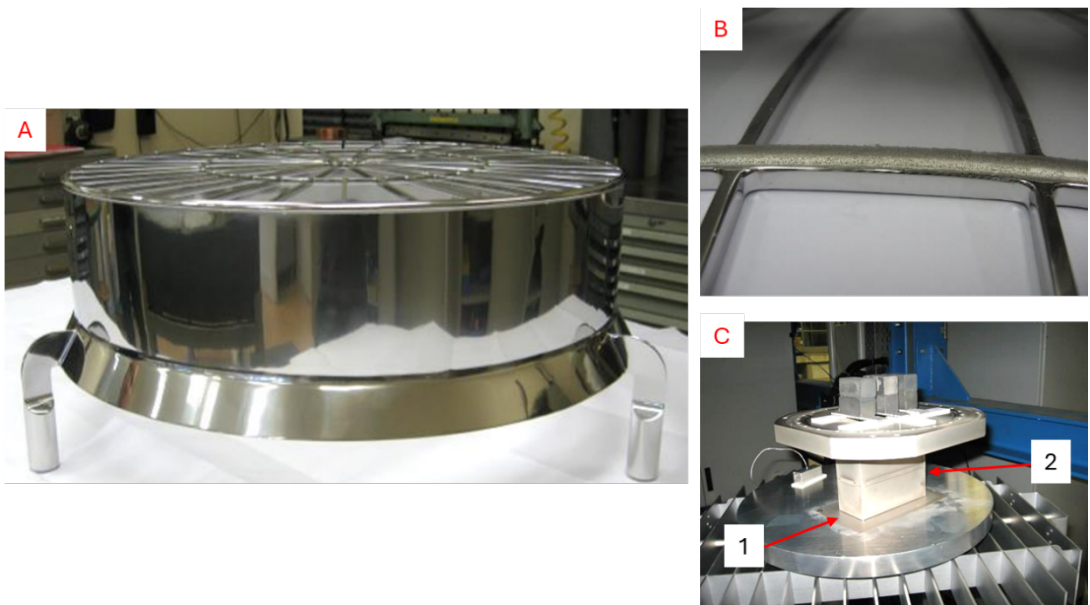

**Fig. 4:** Examples of stainless-steel vacuum brazing. A & B: CLOUD fan hood. C: Roman pot.

Vacuum brazing is mainly recognized for its ability to join dissimilar materials. Figure 5 shows four examples: A: A niobium tube in a stainless-steel flange; Pure copper is used as the brazing alloy. Wetting is very good on both materials, but the thermal cycle is critical and must be carefully controlled to avoid formation intermetallics. B: A stainless-steel capillary on molybdenum which is a metal difficult to braze This assembly was made on a small area. C: Copper and titanium tubes. These metals assembled together easily form fragile intermetallics during brazing cycle. The solution shown here is to braze a stainless-steel ring between the central titanium tube and the external copper tubes. D: Glidcop beam and CuNi square tubes. The brazing of Glidcop is only possible with a diffusion barrier made of a thin film. The best solution seems to be the electrodeposition on Glidcop of a first layer made of nickel and a second made of copper.

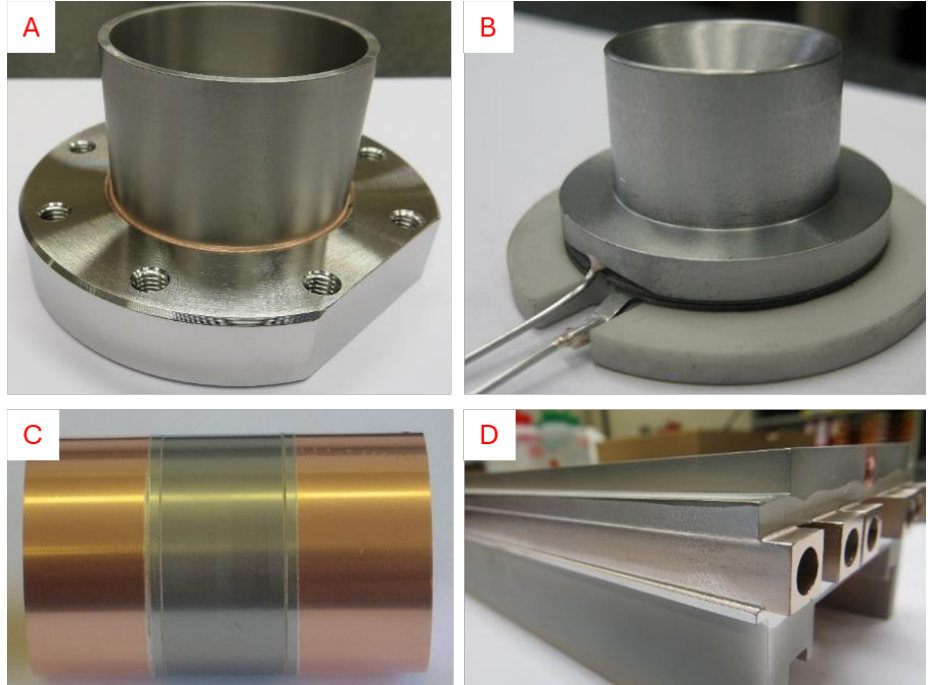

**Fig. 5:** Examples of bi-metallic vacuum brazed junctions: A: niobium/stainless-steel, B: molybdenum/stainless-steel, C: copper/stainless-steel/titanium, D: Glidcop/CuNi.

## 4   Vacuum soldering

As previously discussed, wetting is limited in a vacuum furnace at a temperature corresponding of the melting point of common solder alloys (around 200 °C). Acceptable results can be achieved with pure or electrodeposited silver or copper. Common solders are the two eutectic alloys composed of tin and silver for the first (melting point 221 °C) and tin and lead for the second (melting point 183 °C). These elements have a very low vapour pressure at 200 °C, making them compatible with a soldering in vacuum. However, this only applies if the alloys are of very high purity and do not contain traces of cadmium or zinc, as is often the case with cheap commercial solders.

Vacuum soldering is often chosen when it is necessary to preserve the integrity of the assembled components. One example is the vacuum soldering of the High Temperature Superconducting (HTS) tapes for the LHC current leads, see Ref. [3]. The tapes, which are made of BSCCO superconducting filaments in a silver-gold matrix, are sensitive to high temperature thermal cycle. They may also be damaged if chemical flux is used. They must, however, be assembled to obtain a good thermal conduction on the high superconducting section of the current leads.

Vacuum soldering has been tested and found to be the best solution for both assembling the tapes to form stacks and assembling the stacks onto the stainless-steel cylinders, which, together with two copper blocks that are vacuum brazed at the extremities, form the superconducting section of the leads, see Fig. 6.

Sn-Ag solder was used in the first step, followed by Sn-Pb in the second. Grooves were machined into the stainless-steel cylinder, and were copper-plated before the stacks were soldered.

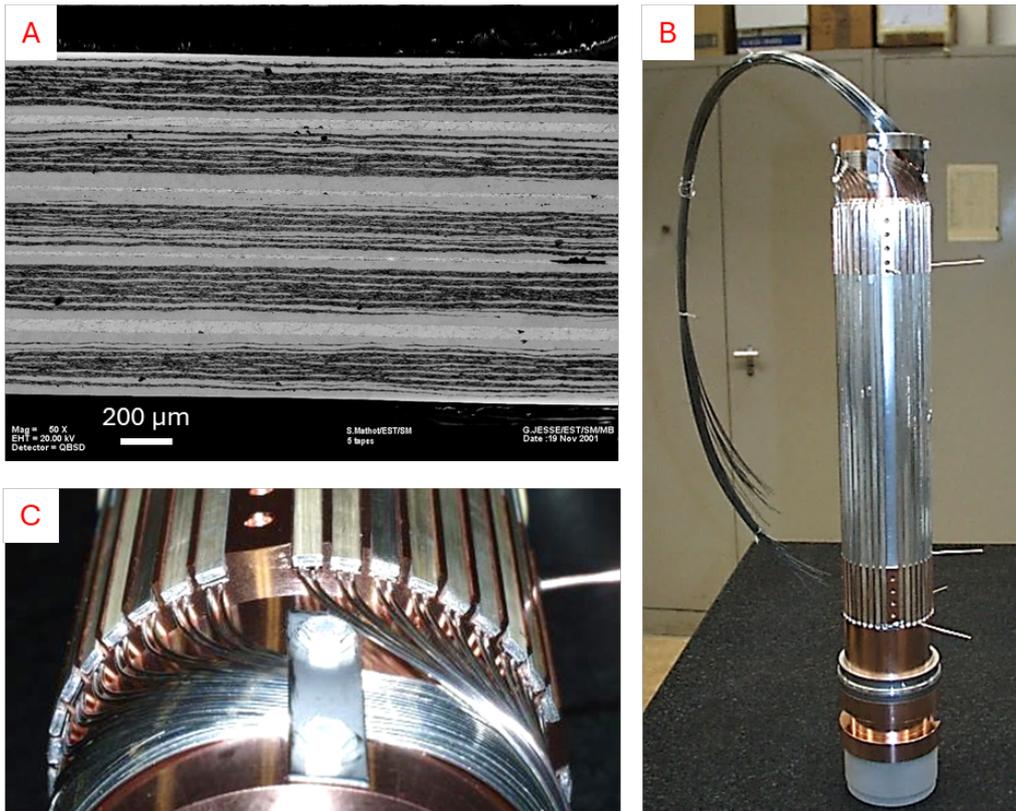

**Fig. 6:** A: Micrography of a vacuum soldered stack. B: Superconducting section of the lead after final soldering. C: Bottom section of the lead with connections with the LTS (Low Temperature Superconducting) wires.

## 5 Vacuum brazing of large components, example with the RFQ's

RFQ's (Radio Frequency Quadrupole) are accelerating cavities made of four vanes that are vacuum brazed together. The two RFQ's manufactured for the Linac4 are the largest assemblies we machined, and vacuum brazed at CERN. Each Linac4 RFQ is composed of 3 modules, each one meter long, with a diameter of about 400 mm and a final weight of 450 kg, see Fig. 7 A and Ref. [4]. Another RFQ model, called HF-RFQ (High-Frequency RFQ) is composed of small modules (500 mm long, with a diameter about 200 mm and a weight of 50 kg) but requires higher precisions in the machining and the assembly, see Fig. 7 B and Ref. [5].

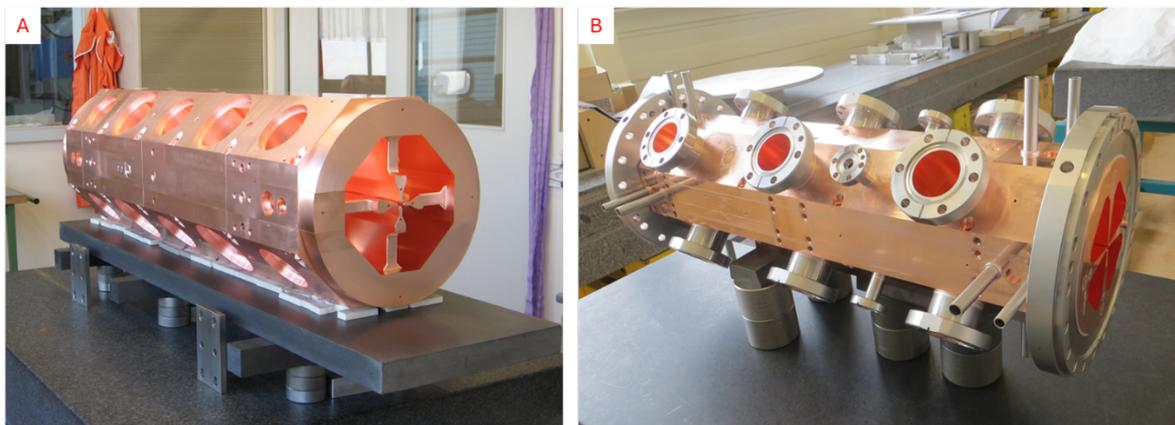

**Fig. 7:** A: Module of the Linac4 RFQ before the first brazing (assembly of the four vanes). B: Module of an HF-RFQ after final brazing (assembly of the flanges on the module).

**Table 2:** RFQ mechanical tolerances

|  | **Linac 4 RFQ** *(325 MHz)* | **HF-RFQ** *(750 MHz)* |
|---|---|---|
| Shape of the vane tip | ± 10 μm | ± 5 μm |
| Position of the vane tip | ± 30 μm | ± 15 μm |
| Shape of the cavity | ± 20 μm | ± 10 μm |
| Vane max. rotation (X or Y) | ± 50 μm | ± 25 μm |
| Vane max. displacement (beam axis) | ± 50 μm | ± 20 μm |
| Gap between two modules | ± 15 μm | ± 10 μm |

Table 2 shows the mechanical tolerances imposed for the two RFQ models, see Ref. [6]. As shown in Fig. 7, the assembly procedure consists of two vacuum brazing steps, the first for the assembling of the vanes in a horizontal position and the second for the assembling the flanges in a vertical position. Maintaining the vane position after a brazing heat cycle at around 800 °C is the main challenge. Machining the individual vanes with high precision is feasible with a high precision machine and a high skills operator (both are required). However, the higher challenge lies in maintaining the precision obtained after the brazing cycle.

We have demonstrated that a major vane of a Linac4-type RFQ that is machined without taking intermediate precautions will deform to an unacceptable level for a vacuum brazing after a vacuum heat treatment at 800 °C. Our procedure for assembling the RFQ involves alternating machining and heat treatments during the production of the individual vanes, thereby reducing the stress induced during the machining phases. During the final machining step, we removed only 150 μm of material from the more sensitive regions of the vanes, i.e. the vane tips and the brazing surfaces. This reduces the level of deformation during the first brazing cycle. Tooling is only used to reduce possible movement during this step and is ineffective in the event of deformation due to stress relief. The thermal brazing cycle is also adapted to the size of the pieces, with slow heating and cooling ramps.

Figure 8 illustrates the preparation phases prior the first brazing on an HF-RFQ: A: Preparation of the individual vanes and placement of the brazing alloy in the grooves. B: Installation of the bottom major vane on the support, C: Installation of the two horizontal vanes (minor vanes). D: The module with the four vanes in place. E: Control of the positioning. F: Installation of the tooling.

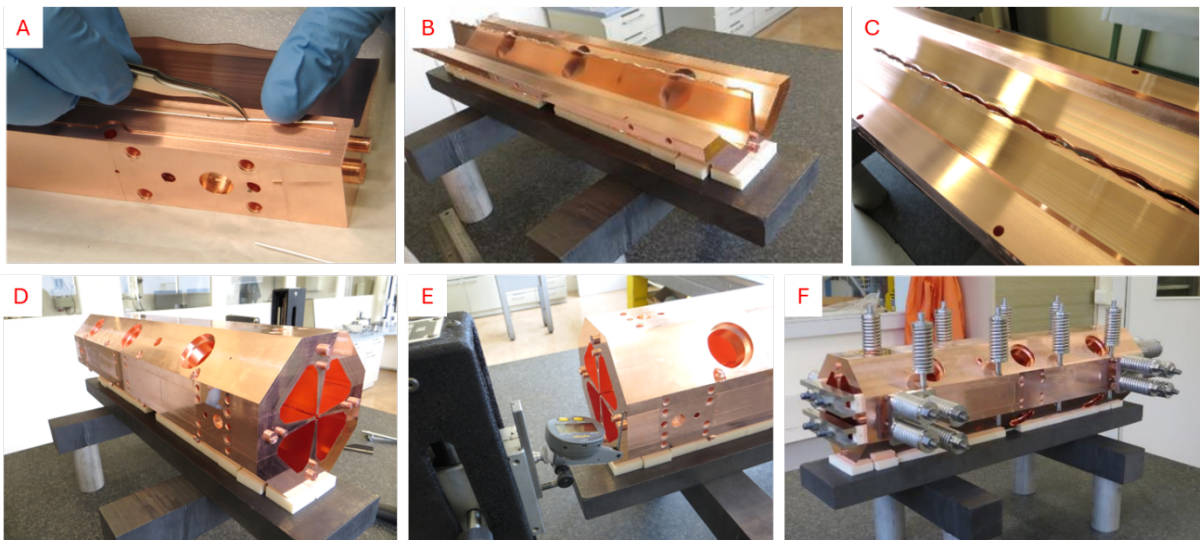

**Fig. 8:** Preparation phases before the first brazing step of an HF-RFQ.

A visual inspection after the first brazing is essential. The brazing alloy must be clearly visible at the interfaces between the brazing surfaces of the vanes, without interruption but also without any excessive flow onto the surfaces. This is particularly important for the internal surfaces of the cavities, see Fig. 9. The white arrows indicate the correct results after first brazing.

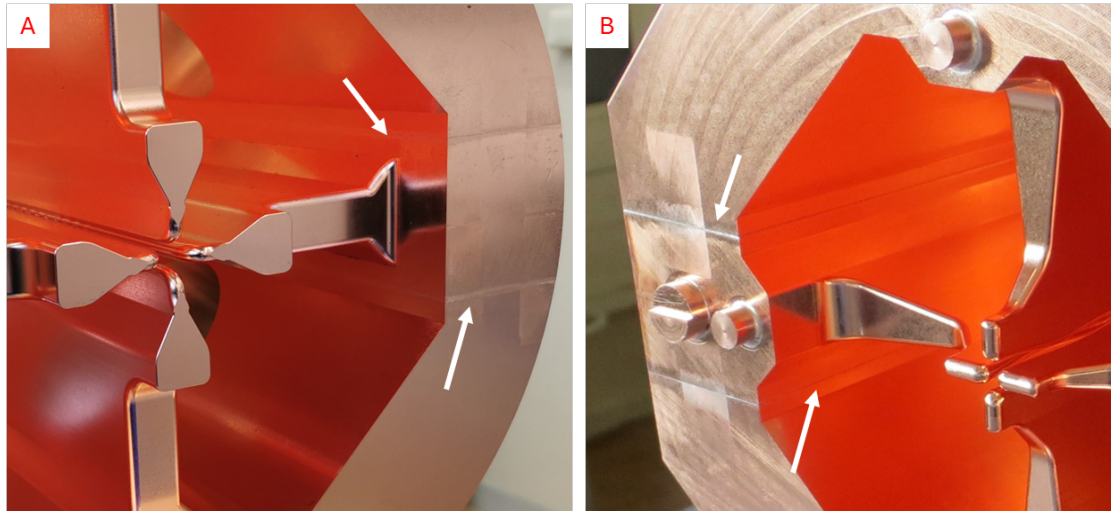

**Fig. 9:** Linac4 RFQ (A) and HF-RFQ (B) modules after first brazing.

The second brazing step is performed in a vertical position after the diameters where flanges must be brazed have been re-machined (finished). Figure 10 shows two examples of RFQ modules inside the vacuum furnace during this step. Tooling is also used to hold the flanges in place.

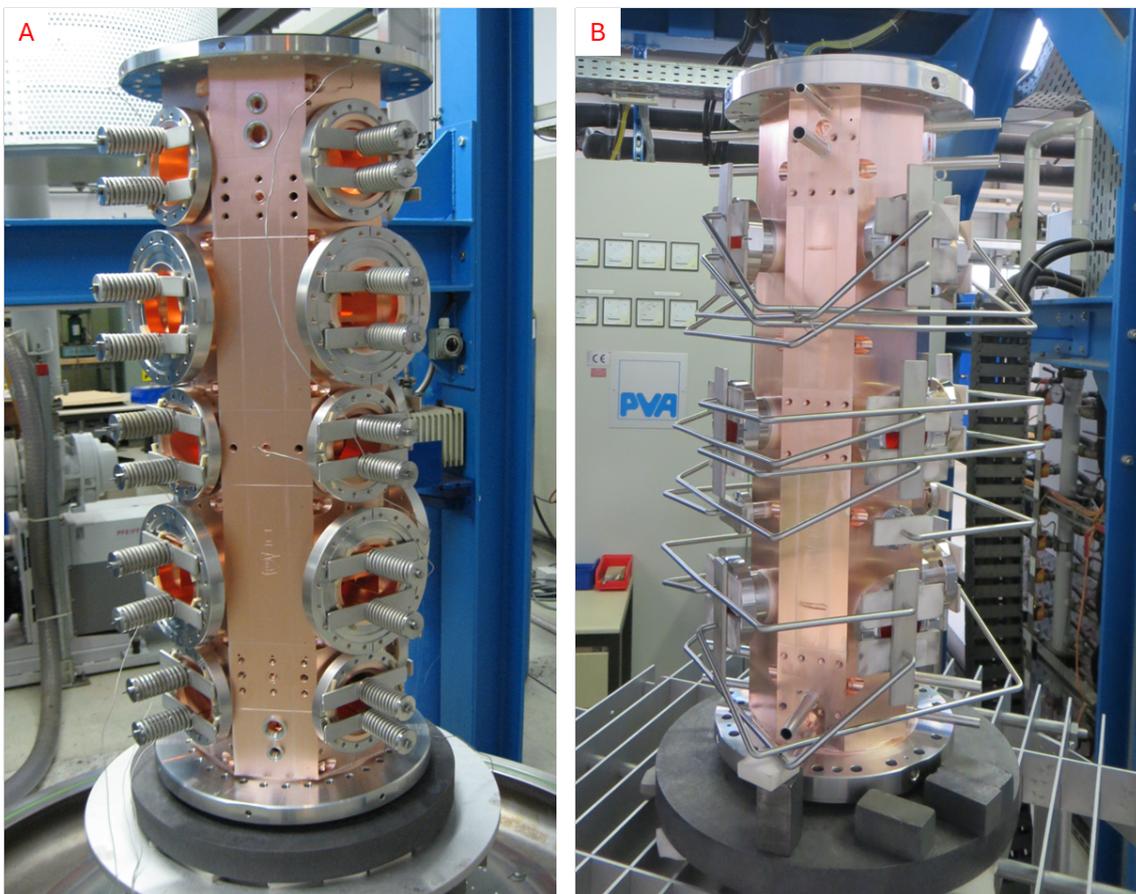

**Fig. 10:** Linac4 RFQ (A) and HF-RFQ (B) modules inside the vacuum furnace for the second brazing step.

Success in assembly, given difficulties such as those encountered in the RFQ's, mainly comes from discussions during the design preparation stage. Discussions after the design has been finalised and the technical drawings have been completed are too late. At CERN, RFQ's are a good example of how all the relevant parties (beam dynamics, RF, construction and design) are involved from the outset of the project, while there is still scope for suggestions about the design!

Specifically, for both the Linac4 and the HF RFQ's, the presence of a fairly sharp angle on either side of the brazing planes between two vanes was requested (see Fig. 11). This allows the progression of the brazing alloy towards the tip of the vanes, in the centre of the cavity, where the electric fields are strongest, to be limited in the event of excessive flow of the brazing alloy. It is absolutely necessary for there to be no brazing material in this area! The RF design has been adapted to make this possible!

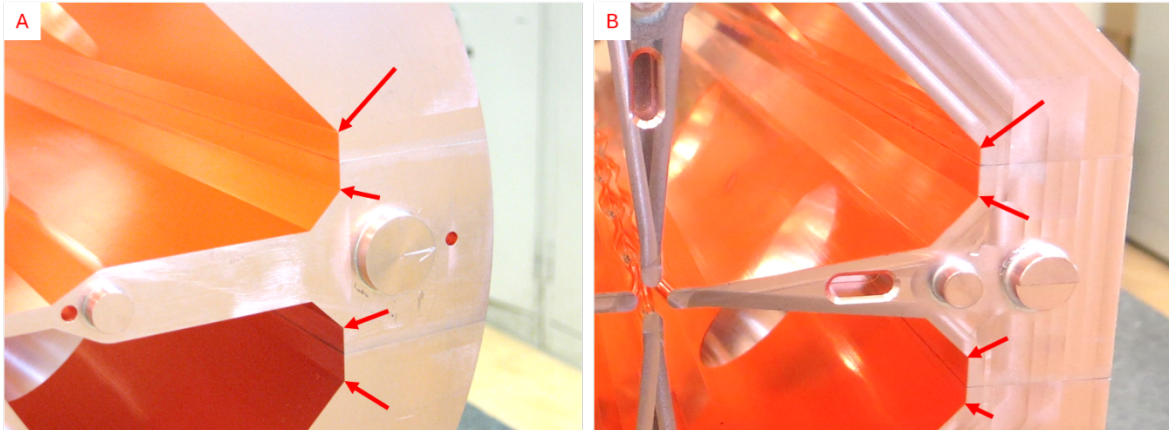

**Fig. 11:** Linac4 RFQ (A) and HF-RFQ (B) modules. Red arrows point the shape of the internal cavity where edges are present of both sides of the brazed interfaces.

## 6    Vacuum brazing of metallized alumina

Vacuum brazing is a unique technique that allows alumina (a widely used insulator) and metal to be assembled. This is important for manufacturing insulating transitions that are compatible with high vacuum and that are required for many components of particle accelerators and detectors.

As the variations in composition, grain size and manufacturing methods can be significant, it is possible to talk about alumina's rather than alumina. Except for sapphire, alumina's are obtained by sintering aluminium oxide ($Al_2O_3$) grains. Adding glass (a mixture of $SiO_2$, CaO, MgO, $ZrO_2$, etc.) reduces the sintering temperature. Technical alumina's, which are used for high vacuum applications, contain between 85 % and 99.9 % of $Al_2O_3$.

The process of metallising alumina using refractory metals was invented in the 1930's. This process, known as the Moly-Manganese process due to its main components, has undergone numerous developments. In 1950, a process based solely on molybdenum and manganese was developed and quickly commercialised. These developments paved the way for the large-scale production of sealed ceramic/metal joints for vacuum applications. Vacuum brazing is the most widely used assembly method for joining metals onto metallised alumina, see Ref. [7].

The Mo-Mn metallisation process uses molybdenum and manganese powders. The powders are mixed with an organic binder and applied to the ceramic parts to be metallised. The resulting layer is approximately 25–30 μm thick.

The trick is then to heat the entire assembly to a high temperature in a carefully selected atmosphere, ensuring that the molybdenum (Mo) remains metallic while the manganese (Mn) oxidises to form manganese oxide (MnO). These conditions can be achieved in a humid hydrogen atmosphere. The metal/oxide equilibrium diagram under a hydrogen atmosphere as a function of partial water pressure shows that, if PH$_2$O > $10^{-1}$ Torr, the MnO oxide is stable at 1200 °C, whereas the MoO$_2$ is unstable and the metal is reduced. Therefore, above 1200 °C in a humid hydrogen atmosphere, the metallisation layer consists of metal powder and MnO. At this temperature, the MnO reacts rapidly with the glassy components of the ceramic (e.g. SiO$_2$, CaO) and even with Al$_2$O$_3$ to form a liquid phase. Upon cooling, this glassy phase solidifies and adheres strongly to the ceramic, also sintering the metal powder. The resulting layer is 12 to 20 μm thick.

Finally, a second layer, usually nickel, is chemically deposited on top of this first layer and sintered at 1000 °C under vacuum. The purpose of this layer is to improve the wetting of the brazing alloy used to assemble the ceramic.

Altough metallisation solves the problem of brazing alloy wetting for ceramic/metal assemblies, the choice of metal must be made with caution. The coefficient of expansion is not the only parameter to consider. A material with a high coefficient of expansion that deforms easily (i.e. with a low elastic limit) will cause less stress in the ceramic than a material with the same coefficient of expansion but a higher elastic limit. A factor known as the Thermomechanical Compatibility Factor (TCF) can be defined as:

$$TCF \cong \frac{\varepsilon_y}{\varepsilon_t} \frac{1}{\sigma_y}$$

Where $\varepsilon_y$ is the metal elastic elongation at brazing temperature, $\varepsilon_t$ is the difference in elongation between metal and ceramic at brazing temperature and $\sigma_y$ is elastic limit of the metal.

The higher this factor is, the easier it is to braze the corresponding metal onto the ceramic. Table 3 shows the calculated TCF values for various metals for the brazing onto alumina at 780°C. We can see that the 'ideal' metal is niobium. Copper is also easy to braze, while stainless steel and tungsten should be avoided. A metal that is often brazed onto alumina is Kovar (an Fe-28Ni-18Co alloy). As can be see, it is indeed the metal with the highest TCF that can be welded to stainless steel. This is an important advantage for many applications.

**Table 3:** TCF factor for different alloys calculated for alumina brazing at 780 °C, see Ref. [8].

| Alloys | TCF |
|---|---|
| Niobium | 88 |
| Platinum | 33 |
| Tantalum | 28 |
| Copper | 20 |
| Titanium | 8.8 |
| Kovar | 7.7 |
| Nickel | 6.7 |
| CuNi | 4.8 |
| Fe-42Ni | 4.5 |
| Monel | 4.0 |
| Invar | 3.7 |
| Molybdenum | 3.5 |
| Stainless steel 304 | 2.9 |
| Inconel 600 | 2.1 |
| Tungsten | 2.0 |

Figure 12 shows some examples of metal/metallized alumina vacuum brazing assemblies: A: Large alumina insulator with Kovar, B: Typical alumina ring with Kovar collars designed to be brazed onto stainless steel flanges, C: Small insulator tube on copper tubes, D: Large alumina disk on copper, E: Copper/alumina and F: Titanium flange/alumina ring/copper tube. Note on pictures C & D the use of molybdenum wires which deform copper during the brazing cycle to insure a continuous contact with the metallized alumina.

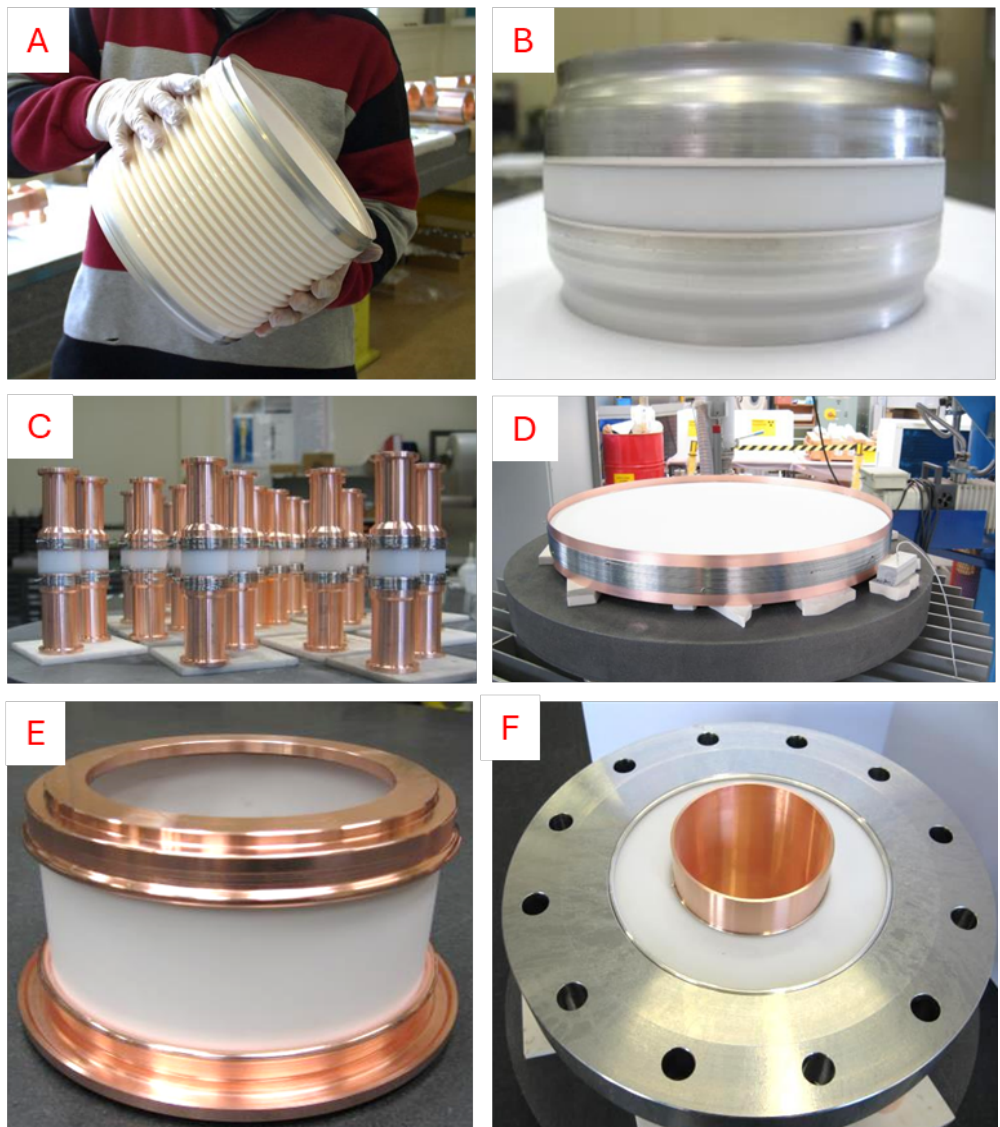

**Fig. 12:** Examples of metal/metallized alumina vacuum brazing assemblies.

## 7  Assembly with active brazing alloys

While most brazing alloys do not wet ceramics, there are certain metals that can react with them, particularly alumina. These metals are reducing metals such as Ti, Zr, Be, etc., which are known as active metals. In the 1960s and 1970s, brazing alloys containing a certain percentage of active metal were developed. A whole series of these alloys have been and are still being developed to enable ceramics to be brazed at different temperatures and on different metals.

However, the development of these 'active' brazing alloys is not straightforward due to the formation of brittle compounds. A AgCuTi-type alloy, based on the AgCu eutectic, which has a brazing temperature suitable for many applications and good wetting on metals, must contain a sufficient percentage of Ti to be active, but not so much to limit the formation of brittle Ti/Cu phases. The optimal composition in this case is around 1.5 % Ti and must be less than 4 %.

Brazing with active alloys is based on chemical reactions between the active metal and alumina. These reactions are complex, and at the Ti/Al2O3 interface, phases such as $Ti_3Al$ and $TiAlO_x$ are present. It is these intermetallic phases that promote wetting by the brazing alloy. It should be noted that this type of brazing is only possible under high vacuum conditions, below $1 \cdot 10^{-5}$ Torr.

It should also be noted that wetting is generally poor: the brazing alloy will adhere to the alumina surface but will not actually flow over it as it would on a metallised surface. Therefore, brazing of a collar with a cover, for example, will be difficult unless the alloy is deposited between them. In active brazing, the joint configuration must take this constraint into account and, in general, brazing on a flat surface without interlocking is more suitable. Therefore, although active brazing eliminates the time-consuming metallisation step, it is generally considered to be more difficult to implement for alumina assembly.

Nevertheless, active brazing has the great advantage of being able to be used for assembling of other ceramics whereas the Mo-Mn metallisation is strictly limited to technical alumina's. Indeed, Ti or Zr, for example, can also react with other oxides, carbides or nitrides. Thus, active brazing alloys adhere to pure alumina (sapphire), carbides such as SiC, graphite or diamond, nitrides such as AlN, etc. Through complex diffusion and reaction mechanisms, it is considered that, in principle, all ceramics can be brazed with an active brazing alloy.

However, to achieve ceramic/metal assembly, the stresses that will be produced during cooling must be considered. Ceramics such as SiC or ferrites, for example, are very fragile and breakages are often observed at the interface if sufficient care is not taken regarding the choice of metal, joint configuration, thermal cycle, etc.

Figure 13 shows examples of metal/ceramics assemblies obtained by active vacuum brazing: A: Large sapphire window (Ø 115 mm) on a niobium collar, alumina rings and Kovar electrodes, B: Metallography of a carbon/CuNi interface, active brazing alloy is CuAgTi, C: AlN/Kovar, D: Small diamond window (Ø 5 mm) on titanium, E: Technical alumina on Monel, active brazing is used in this example to avoid Mo-Mn metallisation, which can only be carried out by (a few) specialised companies, F: $ZrO_2$ tubes with titanium flanges. The figure also includes example of the use of an active brazing alloy for the assembly of metals difficult to braze with conventional alloys, G: Tantalum tube on a stainless-steel flange, H: Tungsten rod in copper.

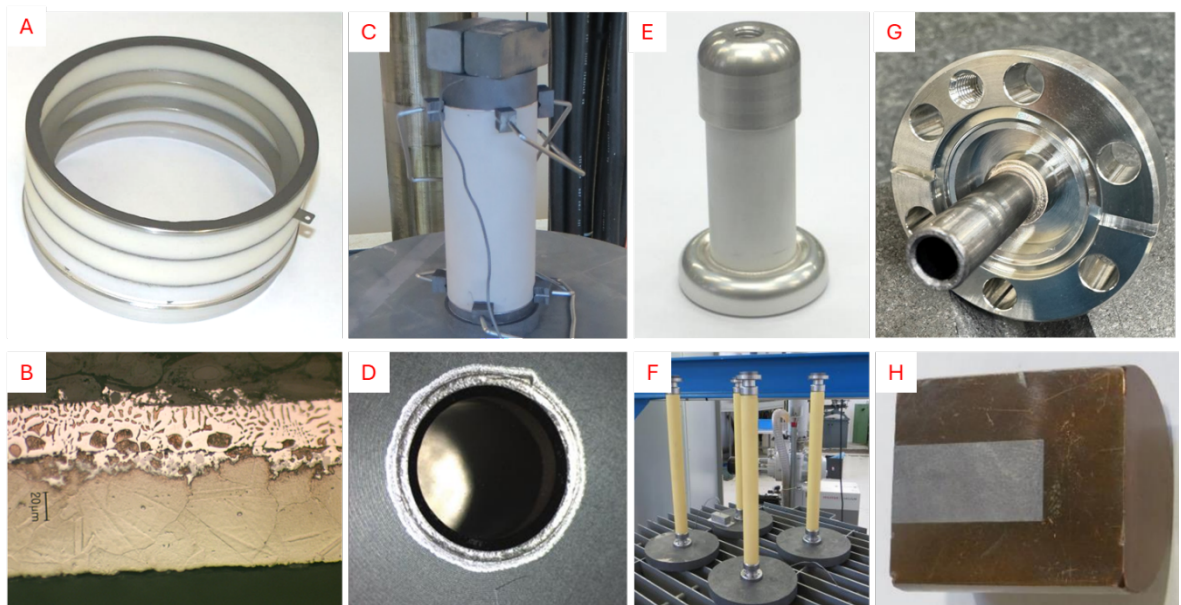

**Fig. 13:** Examples active vacuum brazing assemblies.

# 8 Conclusions

Vacuum brazing is an important assembly technology with key advantages for precision and for joining dissimilar materials. It is also the only technique for joining ceramics and metals in high vacuum applications.

However, the heat cycle may alter the mechanical properties of the parts. The machining process is always highly precise, and the preparation before brazing can be complex.

The key to success in vacuum brazing is to start with a discussion of the design and then follow a strict procedure.